\documentclass{article}

\usepackage[main, final]{neurips_2026_vericode}

\usepackage[utf8]{inputenc} 
\usepackage[T1]{fontenc}    
\usepackage{hyperref}       
\usepackage{url}            
\usepackage{booktabs}       
\usepackage{amsfonts}       
\usepackage{nicefrac}       
\usepackage{microtype}      
\usepackage{xcolor}         
\usepackage{placeins}

\usepackage{newfloat}
\usepackage{listings}
\usepackage{booktabs}

\usepackage{tabularx}
\usepackage{pifont}         
\usepackage{xspace}

\newcommand{\FVSpec}{FVSpec\xspace}

\newcommand{\Secl}[1]{\label{sec:#1}}
\newcommand{\Secr}[1]{Section~\ref{sec:#1}}

\usepackage{tikz}
\usetikzlibrary{arrows.meta,positioning,fit,calc,backgrounds}
\usepackage{adjustbox}
\usepackage{enumitem}

\definecolor{darkorange}{RGB}{204,102,0}

\newsavebox{\fvspecPyImplBox}
\newsavebox{\fvspecPyPbtBox}
\newsavebox{\fvspecImplBox}
\newsavebox{\fvspecSpecBox}

\definecolor{lstKeyword}{RGB}{0,0,170}     
\definecolor{lstTactic}{RGB}{160,40,40}    
\definecolor{lstType}{RGB}{20,120,140}     
\definecolor{lstComment}{RGB}{106,153,85}  
\definecolor{lstString}{RGB}{166,38,164}   
\definecolor{lstBuiltin}{RGB}{0,134,179}   

\lstdefinelanguage{Lean}{
  morekeywords={
    import, open, namespace, end, section, variable, variables, universe, universes,
    structure, class, instance, inductive, coinductive, abbrev, def, theorem, lemma,
    example, axiom, constant, opaque, mutual, deriving, where, attribute, syntax,
    macro, elab, notation, infix, infixl, infixr, postfix, prefix,
    private, protected, noncomputable, partial, unsafe, scoped, local,
    let, fun, if, then, else, match, with, do, return, as, in, at, of, from
  },
  morekeywords=[2]{
    intro, intros, exact, apply, refine, simp, rfl, rw, rewrite, assumption,
    cases, induction, constructor, contradiction, decide, sorry, admit,
    trivial, ring, linarith, omega, norm_num, tauto, exists, use, calc,
    show, have, suffices, obtain, rintro, ext, by
  },
  morekeywords=[3]{
    Type, Prop, Sort, Bool, Nat, Int, String, Char, List, Array, Float, Unit,
    Option, Sum, Prod, And, Or, Not, True, False
  },
  sensitive=true,
  morecomment=[l]{--},
  morecomment=[s]{/-}{-/},
  morestring=[b]{"},
  literate=
    {->}{{$\rightarrow$}}2 {<-}{{$\leftarrow$}}2 {=>}{{$\Rightarrow$}}2
    {→}{{$\rightarrow$}}1 {←}{{$\leftarrow$}}1 {↦}{{$\mapsto$}}1
    {≤}{{$\leq$}}1 {≥}{{$\geq$}}1 {≠}{{$\neq$}}1 {≡}{{$\equiv$}}1
    {∀}{{$\forall$}}1 {∃}{{$\exists$}}1 {¬}{{$\neg$}}1
    {∧}{{$\wedge$}}1 {∨}{{$\vee$}}1 {⟨}{{$\langle$}}1 {⟩}{{$\rangle$}}1
    {·}{{$\cdot$}}1 {ℕ}{{$\mathbb{N}$}}1 {ℤ}{{$\mathbb{Z}$}}1
}

\lstdefinestyle{code}{
  basicstyle=\ttfamily\footnotesize,
  keywordstyle=\color{lstKeyword}\bfseries,
  keywordstyle=[2]\color{lstTactic},
  keywordstyle=[3]\color{lstType}\bfseries,
  commentstyle=\color{lstComment}\itshape,
  stringstyle=\color{lstString},
  columns=fullflexible,
  keepspaces=true,
  showstringspaces=false,
  frame=none,
  breaklines=true,
  breakatwhitespace=true,
  numbers=none,
  emph={[1]hypothesis,strategies,bisect,given,insort,sorted,assert,
        st,pytest,np,numpy,torch,Hypothesis},
  emphstyle={[1]\color{lstBuiltin}}
}

\title{\FVSpec: Real-World Property-Based Tests as Lean Challenges}

\author{
  Quinn Dougherty\thanks{Forall R\&D, work completed at Galois Inc, correspondence to \texttt{quinn@for-all.dev}},
  Max von Hippel\thanks{Some work completed at Benchify},
  Simon Henniger\thanks{Harvard University},
  Hazel Shackleton\thanks{Galois Inc},
  Mike Dodds\footnotemark[4]
}
\date{}

\begin{document}

\maketitle
\setcounter{footnote}{0}

\begin{abstract}
As AI systems generate an ever-growing share of the world's code, formal verification offers a principled way to ensure that code is correct, and AI itself may be able to shoulder much of the verification burden. Yet we lack a clear picture of how well today's models and agents perform on verification tasks drawn from real-world software, since existing benchmarks rely largely on curated or synthetic problems, or focus on math rather than program verification. To close this gap, we present a benchmark for evaluating AI on real-world formal software verification tasks. We first scrape 7,413 distinct property-based tests (PBTs) from real-world Python repositories, then automatically translate 2,623 of them (35\%) into 9,415 Lean 4 specifications with \texttt{sorry} placeholders ($\approx$3.5 formalizations/PBT).
Translating PBTs to Lean specs is challenging: it requires modeling Python in Lean, inferring the property encoded in a PBT, and handling side effects. We describe an agentic  pipeline for transpiling PBTs into Lean, evaluate coverage and quality metrics, and provide baselines for proof generation using several approaches. All code (scraper and agents) and data (PBTs and Lean specs) are open source.
\end{abstract}
\section{Introduction}
\Secl{intro}

Several AI safety proposals are premised on implementing safeguards by way of 
\emph{formal verification} 
(FV)\footnote{FV forms a subset of \emph{formal methods}, which also include PBT, model checking, and abstract interpretation.} 
-- 
that is, software systems that can produce and 
certify proofs~\citep{safeguardedAI,dalrymple2024guaranteed,sfo,sfo2}.
Software verification is therefore a key goal for AI safety, yet, 
currently, there is little evidence that AI can automate verification.
To achieve this, we would first need high-quality benchmarks on which to train.
Existing benchmarks 
are too small (e.g.~\citet{loughridge2024dafnybench},~\citet{ravi}),
focused on math rather than software verification (e.g.~\citet{glazer2024frontiermath}),
or derived from expert-driven verification projects (as in~\citet{chakraborty2024neural}) that 
do not reflect the tasks involved in a typical software project:
bootstrapping verification in real-world software is much more challenging 
than verifying software which was built to be verified from day one. 
For instance, the former, unlike the latter, often requires reasoning about black-box external systems and 
the interactions of several different components, 
potentially in different languages 
and architectures~\citep{buro2020equational,perconti2014verifying}.

To reflect the needs of real-world verification, we propose a new benchmark, \FVSpec, where we derive verification challenges from publicly available property-based tests (PBTs). PBT is a ``close sibling'' to FV which has seen significant adoption, sometimes called ``lightweight formal methods''~\citep{JacksonWing1996}. In both FV and PBT, engineers write \emph{specifications}: logical properties that the software must obey. For example, we might require that a Python function \texttt{isort(lst)} always returns a sorted permutation of its input \texttt{lst}. The difference lies in how this property is checked. In FV, the property is proved using mathematical techniques, resulting in almost-perfect confidence, but requiring proof engineering by highly expert teams~\citep{dodds2022formally} (or perhaps, in the future, agents). In PBT, tests are randomly generated based on general rules that look like theorems. 

First, we scrape 7{,}413 distinct PBTs written in Python's Hypothesis framework~\citep{maciver2019hypothesis}, creating \FVSpec:PBT.
Each PBT can be seen as a conjecture about a piece of code, which the author would like to prove or disprove. We translate code and PBTs to the prover Lean, which is becoming the de-facto standard in AI verification. In the process of translation, each PBT becomes a \textit{conjecture}.
We refer to our resultant corpus of conjectures and code as \FVSpec:FV.
Once the Lean version of the conjecture is proven, it becomes a \emph{theorem}.
A similar approach was previously used by \citet{dougherty2025proving} on FVAPPS. In FVAPPS, source programs were taken from the APPS benchmark \citep{hendrycks2021apps}, then specs were inferred by an LLM and translated to Hypothesis PBTs, and then lifted to theorems. We apply the same process to PBTs found in the wild.

Our main contributions are:
\begin{enumerate}[leftmargin=1.5em]
\item \textbf{\FVSpec:PBT} — 7{,}413 deduplicated Python PBTs scraped from 497 open-source repositories, the first large PBT dataset drawn from ordinary software development\footnote{Released on HuggingFace in January 2026, before the Hypothesis Corpus~\cite{devoe2026hypothesis}; see Appendix~\ref{sec:dataset-hc}.}.
\item \textbf{\FVSpec:FV} — 9{,}415 Lean~4 FV challenge samples lifted from 2{,}623 of those PBTs by an agentic transpilation pipeline, with detailed metadata (8 FV challenges per PBT on average).
\item \textbf{Baseline evaluations} of Claude Sonnet~4.6, Claude Opus~4.7, and GPT~5.4 showing mean 70\% success on easies and mean 49\% on hards.
\item \textbf{Open-source release} of the full pipeline\footnote{(consisting of scraper, dependency extractor, agent, and post-production scripts)} and datasets under MIT/Apache dual license.
\end{enumerate}

\begin{figure*}[t]
\centering
\begin{adjustbox}{width=0.75\textwidth}
\begin{tikzpicture}[
  font=\footnotesize,
  box/.style={draw, rounded corners=2pt, fill=gray!5, inner sep=4pt, align=center,
              text width=22mm, minimum height=11mm, line width=0.4pt},
  outbox/.style={draw, rounded corners=2pt, fill=blue!8, inner sep=4pt, align=center,
              text width=22mm, minimum height=11mm, line width=0.6pt,
              font=\footnotesize\bfseries},
  arr/.style={-{Latex[length=2mm]}, gray!50!black, line width=0.5pt},
  group/.style={draw, dashed, rounded corners, inner sep=4mm,
                gray!60!black, line width=0.6pt},
  hdr/.style={font=\small\sffamily\bfseries},
  source/.style={font=\footnotesize\itshape},
  node distance=4mm and 4mm,
]
  \node[source]                 (gh)    {GitHub};
  \node[box, right=8mm of gh]   (disc1) {Discovery\\\scriptsize\rmfamily 18{,}494 repos};
  \node[box, right=of disc1]    (lic)   {License filter\\\scriptsize\rmfamily copyleft~$\to$~drop};
  \node[box, right=of lic]      (ext)   {Extract\\\scriptsize\rmfamily \texttt{@given} via ripgrep};
  \node[outbox, right=of ext]   (pbtout){7{,}413 PBTs\\\scriptsize\rmfamily\mdseries 497 repos};
  \draw[arr] (gh.east) -- (disc1.west);
  \foreach \a/\b in {disc1/lic, lic/ext, ext/pbtout} \draw[arr] (\a) -- (\b);

  \node[box, below=16mm of disc1] (disc2) {Function discovery\\\scriptsize\rmfamily tree-sitter};
  \node[box, right=of disc2]      (form)  {Agentic transpilation\\\scriptsize\rmfamily LSP repair loop};
  \node[box, right=of form]       (post)  {Post-production\\\scriptsize\rmfamily merge / validate};
  \node[outbox, right=of post]    (fvout) {9{,}415 samples\\\scriptsize\rmfamily\mdseries 2{,}623 PBTs};
  \foreach \a/\b in {disc2/form, form/post, post/fvout} \draw[arr] (\a) -- (\b);

  \draw[arr] ([xshift=-2.5mm]form.north)
        .. controls ([xshift=-2.5mm,yshift=6mm]form.north)
                and ([xshift= 2.5mm,yshift=6mm]form.north)
        .. ([xshift= 2.5mm]form.north);

  \node[group, fit=(disc1)(pbtout)] (g1) {};
  \node[group, fit=(disc2)(fvout)]  (g2) {};

  \node[hdr, anchor=south west, xshift=-2cm, yshift=0.3mm] at (g1.north east) {\FVSpec:PBT};
  \node[hdr, anchor=south west, xshift=-2cm,  yshift=0.3mm] at (g2.north east) {\FVSpec:FV};

  \draw[arr] ([xshift=-0.5cm] pbtout.south) -- ([xshift=-0.5cm,yshift=-8mm]pbtout.south) -| (disc2.north);
\end{tikzpicture}
\end{adjustbox}
\caption{The full \FVSpec generation pipeline (FVC = FV challenge). \emph{Top} (\FVSpec:PBT):
the input PBT corpus is scraped from public repos. \emph{Bottom} (\FVSpec:FV): each PBT is lifted to a
Lean (Impl, Spec) pair via tree-sitter--based function discovery, an
agentic transpilation step (with LSP-repair), and a
post-production step that validates, deduplicates, and
grades the output; 7{,}413 PBTs enter and 9{,}415 FVCs (from 2{,}623 distinct PBTs) emerge.}
\label{fig:pipeline}
\end{figure*}
\section{Task Definition}
\Secl{task}

Each problem 
consists of four components: (1) the original Python
function under test and (2) PBT testing it, (3) a Lean implementation of the same function, and (4) a Lean theorem corresponding to the PBT, with a \texttt{sorry} placeholder where the proof should go. The AI's task is to fill in every \texttt{sorry} with a Lean proof (or a disproof) that compiles cleanly under \texttt{lake build}; submissions are then scored by whether the proof retains any \texttt{sorry} or special-purpose axioms. Each problem also includes a URL pointing to the original code and license information.
As a running example, consider \texttt{isort(lst)}, illustrated in Figure~\ref{fig:running-example}.
We encounter two issues when translating PBTs into Lean.

\begin{lrbox}{\fvspecPyImplBox}%
\begin{minipage}{0.36\linewidth}
{\small\sffamily\textbf{Python: implementation of \texttt{isort}}}\par\smallskip
\begin{lstlisting}[style=code, language=Python, basicstyle=\ttfamily\scriptsize]
from bisect import insort

def isort(lst):
    out = []
    for x in lst: insort(out, x)
    return out
\end{lstlisting}
\end{minipage}%
\end{lrbox}%

\begin{lrbox}{\fvspecImplBox}%
\begin{minipage}{0.46\linewidth}
{\small\sffamily\textbf{Lean: \texttt{Impl.lean}}}\par\smallskip
\begin{lstlisting}[style=code, language=Lean, basicstyle=\ttfamily\scriptsize]
namespace Fvspec.Impl

def insert (x : Int) : List Int -> List Int
  | []      => [x]
  | y :: ys => if x <= y then x :: y :: ys
               else y :: insert x ys

def isort : List Int -> List Int
  | []      => []
  | x :: xs => insert x (isort xs)

end Fvspec.Impl
\end{lstlisting}
\end{minipage}%
\end{lrbox}%

\begin{lrbox}{\fvspecPyPbtBox}%
\begin{minipage}{0.36\linewidth}
{\small\sffamily\textbf{Python: Hypothesis PBT}}\par\smallskip
\begin{lstlisting}[style=code, language=Python, basicstyle=\ttfamily\scriptsize]
from collections import Counter
from hypothesis import given
import hypothesis.strategies as st
@given(st.lists(st.integers()))
def test_isort(lst):
    out = isort(lst)
    pairs = zip(out, out[1:])
    assert all(a <= b for a, b in pairs)
    assert Counter(out) == Counter(lst)
\end{lstlisting}
\end{minipage}%
\end{lrbox}%

\begin{lrbox}{\fvspecSpecBox}%
\begin{minipage}{0.46\linewidth}
{\small\sffamily\textbf{Lean: \texttt{Spec.lean}}}\par\smallskip
\begin{lstlisting}[style=code, language=Lean, basicstyle=\ttfamily\scriptsize]
import Fvspec.Impl
open Fvspec.Impl

theorem isort_sorted (l : List Int) :
  (isort l).Sorted LE.le := sorry

theorem isort_perm (l : List Int) :
  (isort l).Perm l := sorry
\end{lstlisting}
\end{minipage}%
\end{lrbox}%

\begin{figure}[h]
\centering
\begin{adjustbox}{width=0.75\textwidth}
\centering
\begin{tikzpicture}[
  box/.style={draw, rounded corners, inner sep=4pt, fill=gray!5},
  arr/.style={-{Latex[length=2.5mm]}, thick, gray!50!black},
  group/.style={draw, dashed, rounded corners, inner sep=2mm,
                gray!60!black, line width=0.6pt}
]
  \node[box, anchor=east] (pyimpl)
       at (-0.5cm,  2.2cm) {\usebox{\fvspecPyImplBox}};
  \node[box, anchor=west] (limpl)
       at ( 0.5cm,  2.2cm) {\usebox{\fvspecImplBox}};
  \node[box, anchor=east] (pypbt)
       at (-0.5cm, -2.2cm) {\usebox{\fvspecPyPbtBox}};
  \node[box, anchor=west] (lspec)
       at ( 0.5cm, -2.2cm) {\usebox{\fvspecSpecBox}};
  \node[group, fit=(pyimpl)(pypbt),
        label={[font=\small\sffamily\bfseries]above:{\FVSpec:PBT}}] {};
  \node[group, fit=(limpl)(lspec),
        label={[font=\small\sffamily\bfseries]above:{\FVSpec:FV}}] {};
  \draw[arr] (pyimpl.east) -- (limpl.west);
  \draw[arr] (pypbt.east)  -- (lspec.west);
\end{tikzpicture}
\end{adjustbox}
\caption{Each problem is
partitioned into the \emph{PBT corpus} (\FVSpec:PBT, left)---the original
Python implementation and PBT we scrape from GitHub---and the
\emph{formal-verification challenge} (\FVSpec:FV, right) we transpile from
it: a complete Lean re-implementation and theorem statement
generalizing the PBT. The AI must discharge both Lean \texttt{sorry}
placeholders, one per theorem (or negations).}
\label{fig:running-example}
\end{figure}

\paragraph{Nonsensical theorems pre-validation.} Our pipeline can emit well-formed but useless statements in the following three ways.

\emph{(i) Trivial Unit-typed conjectures.} When the agent cannot recover a function's return type, it falls back on \texttt{Unit}. The resulting theorem is a tautology---e.g.\ \texttt{theorem isort\_check : isort lst = () := by plausible}, which holds for any total function into \texttt{Unit} and so carries no value. The \texttt{Unit} is on the function under test in \texttt{Impl.lean}, and the conjecture is trivial as a result. 

\emph{(ii) Ill-typed or unresolved-symbol theorems.} The agent works from the Python source and writes Lean by analogy. When it invents a Lean name that does not exist---e.g., writing \texttt{theorem isort\_perm (l : List Int) : (isort l).permutationOf l := sorry}, where \texttt{permutationOf} is not a member of \texttt{List}\footnote{(the correct name is \texttt{List.Perm})}---or attaches a type signature inconsistent with the surrounding code, the Lean elaborator rejects the file with errors like \emph{unknown identifier}, \emph{type mismatch}, etc.

\emph{(iii) Spec theorems leaking into \texttt{Impl.lean}.} Our first run used separate agents for \texttt{Impl.lean} (definitions) and \texttt{Spec.lean} (theorems). The implementation agent is explicitly instructed to emit no theorems, but occasionally it pre-empts the spec agent by inserting a \texttt{namespace Fvspec.Spec}\ldots\texttt{theorem isort\_sorted}\ldots \texttt{end Fvspec.Spec} block inside \texttt{Impl.lean}, which violates our file split and would silently  expand the proof obligation if left in.

We catch (i) by regex, drop (ii) at compile time, and strip (iii) with a namespace-aware filter that removes \texttt{namespace Fvspec.Spec} blocks found in \texttt{Impl.lean}. A small residual (0.7\% of samples) carries naked \texttt{theorem} declarations in \texttt{Impl.lean} written outside any namespace; because grading targets only \texttt{Spec.lean}, these do not expand the proof obligation, but they represent stray proof activity that escaped the filter. A theorem that survives all three filters is a reasonable challenge problem even if it has drifted from the original PBT.

\paragraph{Lean as a (nearly) perfect oracle.} The Lean kernel arbitrates correctness, which is qualitatively stronger than statistical sampling. But it is not adversarially perfect: an AI can satisfy the kernel by \emph{changing the question}---adding an axiom, weakening hypotheses, or exploiting a vulnerability in Lean~\citep{gopinathan2026leanbug,vonhippel2026slopless}. To make gaming visible, every submission is scored on two axes: a binary \emph{proved} flag (the file compiles under \texttt{lake build} \emph{and} every original \texttt{sorry} has been discharged) and partial credit (fraction of original \texttt{sorry}s discharged). We additionally record any axiom dependencies the proof introduces, since an AI can trivialize a goal by quietly importing strong axioms; we do not currently auto-fail on extra axioms, but report them so reviewers can flag suspicious solutions. 
Submissions can additionally be re-checked through the independent \texttt{leanprover/comparator}~\cite{leancomparator2026} kernel. 

\section{Methods}

We had to solve two problems: first, how to scrape real-world PBTs, and seecond, how to convert PBTs into FV challenges at scale.

\subsection{Scraping the PBT Corpus}
\Secl{dataset}

We scrape real-world PBTs from public GitHub repositories. To identify which repos are interesting to us, we consider dependencies: if a package depends on Hypothesis~\cite{maciver2019hypothesis}, a widely-used PBT library, it is of interest.  Our scraper operates in three stages.
\emph{(1) Discovery.} We crawl the Hypothesis dependency graph using PyGitHub~\cite{pygithub}.
\emph{(2) Licensing.} We skip repos with an explicitly non-permissive license (e.g.\ AGPL, GPL, LGPL, MPL), and record the license string---including ``unknown'' for repositories with no detectable license---alongside the rest of each repository's metadata. Our thinking here is to avoid software that is licensed in a way that would preempt using our benchmark in nonacademic environments.
\emph{(3) Extraction.} We shallow-clone each accepted repository, locate PBTs via \texttt{ripgrep}, and parse the surrounding Python to recover the transitive closure of locally-defined functions called therein.
A 24-hour run produced 54{,}345 PBTs.

\subsubsection{Deduplication}
\label{sec:dataset-dedup}

Most of those 54{,}345 PBTs are copies. Our first pass deduplicated by repository: we grouped repos by the base name of their URL\footnote{(e.g., \texttt{SunsetMkt/pytorch} and \texttt{njzjz/pytorch} both reduce to \texttt{pytorch})}, kept the most-starred repo in each group, and dropped tests from the others whose whitespace-normalized SHA-256 appeared in the kept one. That left 28{,}073 PBTs across 718 repositories, but only 8{,}314 distinct tests.  Most of the non-distinct PBTs came from PyTorch, whose Hypothesis suite appears, whole or in part, in at least 42 repositories holding 39{,}386 of the 54{,}345 PBTs. Only 24 of them are named \texttt{pytorch}; the rest have unrelated names such as \texttt{amd/ZenDNN-pytorch} or \texttt{fathnd/homomorphic}. The copies come from different PyTorch versions and carry 33 different sets of PBTs, with no PBT appearing in all 42. Other projects are copied too: without the PyTorch copies, the remaining 14{,}959 PBTs contain only 6{,}567 distinct PBTs.  We therefore deduplicate on the PBT alone, keying on a SHA-256 of the PBT with comments and blank lines removed and whitespace collapsed, and keep one canonical copy per key, chosen by stars, then forks, then row id. The resulting corpus has 8{,}314 PBTs from 538 repos owned by 466 GitHub users, spanning 476 project names. The public release of \FVSpec:PBT contains the 21{,}746 name-deduplicated PBTs whose dependencies we could re-extract from a fresh clone, covering 7{,}413 of the distinct tests, and we release code for the PBT-deduplication.

\subsubsection{Diversity}
\label{sec:dataset-diversity}

A perennial concern with scraped benchmarks is concentration: SWE-bench~\cite{jimenez2024swebench}, for instance, draws all 2{,}294 of its tasks from just 12 repositories, with \texttt{django} supplying 37\% and the top three (\texttt{django}, \texttt{sympy}, \texttt{scikit-learn}) supplying 64\%. Our released corpus is far more spread out---it draws from 497 repositories, no single one of which contributes more than 11.5\% of the PBTs and whose top ten together contribute 30.8\%---and most of those repositories are unstarred personal projects rather than curated showcase code. The median PBT is just 10~LoC, but exercises around $2.8\times$ its own length in project code.

To illustrate the diversity of our dataset, consider the result of drawing ten repositories from it uniformly at random:\footnote{Reproduce with \texttt{random.seed(13); random.sample(sorted(r.name for r in repos if r.stars >= 10), 10)} against the 497-repo deduplicated corpus; the star floor filters out personal forks.}
\texttt{flytekit} (workflow SDK),
\texttt{landlab} (earth-surface modelling toolkit),
\texttt{onnx-embedding} (ONNX-backed embedding library),
\texttt{ch-backup} (ClickHouse backup tool),
\texttt{logot} (log-output assertion library),
\texttt{mlstacks} (MLOps infrastructure provisioner),
\texttt{whenever} (datetime library),
\texttt{anls\_star\_metric} (document-QA evaluation metric),
\texttt{jsondiff} (JSON diffing library), and
\texttt{sec-parser} (SEC-filings parser).
This is broadly representative of real-world software engineering, as opposed to academic or competitive programming.

\begin{figure}[t]
  \centering
  \includegraphics[width=\linewidth]{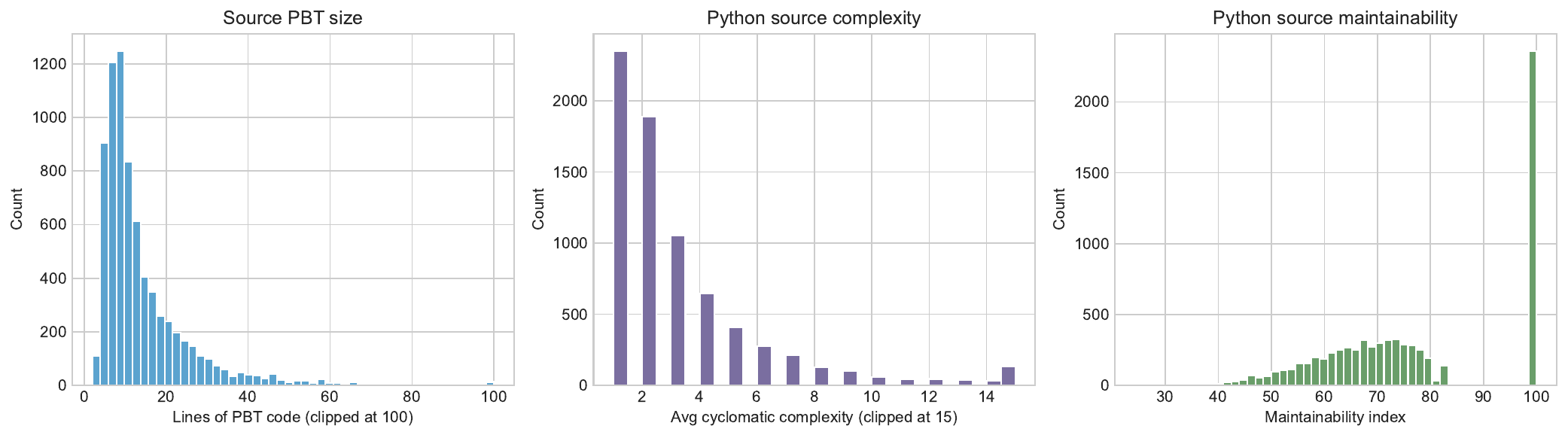}
  \caption{\textbf{Python source characteristics.} Code-quality metrics for the 7{,}413 deduplicated PBTs, computed by Radon~\citep{radon}. \emph{Left}: lines of code.
  \emph{Center}: cyclomatic complexity---the number of linearly independent execution paths through the function, i.e.\ one plus the number of binary branch points~\citep{mccabe1976complexity}; higher values indicate more complex control flow.
  \emph{Right}: maintainability index---a 0--100 composite of Halstead volume~\citep{halstead1977elements}, cyclomatic complexity, and source lines, where higher is more maintainable; the spike at 100 is an artefact of Radon capping the score.}
  \label{fig:python_source}
\end{figure}

Figure~\ref{fig:python_source} summarizes three code-quality metrics across the corpus.
The source tests span a wide range of complexity, from trivial single-assertion tests to multi-strategy tests exercising complex library APIs---diversity inherited directly from real-world engineering practice and not curated for difficulty.

\begin{figure}[t]
  \centering
  \includegraphics[width=\linewidth]{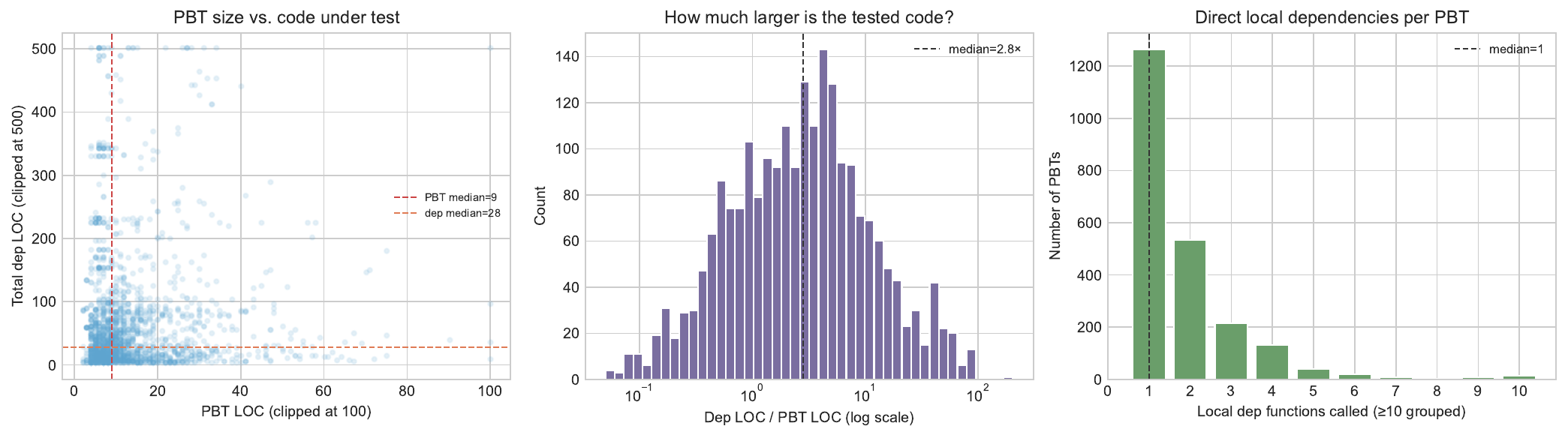}
  \caption{Three views of the relationship between PBT size and the size of the
  code being exercised, over the $n=2{,}238$ post-deduplication PBTs whose
  local dependencies were recoverable.
  (The remaining $5{,}175$ exercise only standard-library or external code, or had dependency extraction fail; the figure is therefore a subset of the full $7{,}413$-PBT corpus.)
  (a)~Per-PBT scatter of PBT LoC against the total LoC of the directly-called
  dependencies.
  (b)~Histogram of the ratio between dependency LoC and PBT LoC (log scale);
  the median PBT exercises code $2.8\times$ its own length.
  (c)~Histogram of the number of direct local function calls per PBT.}
\label{fig:pbt-vs-deps}
\end{figure}

\subsection{Turning PBTs into Lean challenges}
\Secl{pipeline}
We lift each PBT in corpus (\FVSpec:PBT) into a Lean
(\texttt{Impl.lean}, \texttt{Spec.lean}) pair (Figure~\ref{fig:pipeline},
bottom) in three phases: function discovery, agentic transpilation, and
post-production.
We call the resulting challenge corpus \FVSpec:FV.
 The pipeline is modeled on
FVAPPS~\citep{dougherty2025proving}, except it starts from PBTs rather
than programming puzzles, and adds LSP-repair.

\emph{Function discovery.}\label{sec:pipeline-discovery}
For each PBT we use tree-sitter to extract the function under test and the transitive closure of its local dependencies. Larger PBTs pull in tens to hundreds of lines of internal helpers. When discovery fails---typically due
to metaclasses, decorators, or other dynamic dispatch---we emit a stub
 with generic type parameters so the formalization agent still
receives well-formed input.

\emph{Agentic transpilation.} A formalization agent translates each sample, producing both
\texttt{Impl.lean} (a computable port of the function under test and its dependencies, no
\texttt{sorry}) and \texttt{Spec.lean} (one theorem per Hypothesis
assertion, with proofs left as \texttt{sorry}).
After each generation step we type-check the output via the Lean LSP and, on failure, feed the compiler output back to the agent. The loop terminates on success or when a fixed budget runs out. The agents ran on \texttt{claude-sonnet-4-5} in our first run (5{,}979 released samples), which used separate agents for the implementation and the specification, and on \texttt{claude-sonnet-4-6} in our second run (3{,}436 samples).

\emph{Post-processing.} We first extract LLM turn counts from the raw \texttt{.eval} file archives produced by the formalization agent, which runs as an Inspect~AI~\citep{UKAISI_inspect_2024} task. We then collapse runs into a single JSONL, deduplicating by sample id and tie-breaking with a quality score (structural faithfulness, theorem count, unit-tests, implementation success). Next, we re-run
\texttt{lake build} on each retained sample and drop failures, which we classify into nine families: timeouts, non-wellfounded recursion, unknown identifiers, failures to synthesize, type mismatches, syntax errors, missing functions, application errors, and declaration errors. We then run a custom AST analyzer on the Lean artifacts to extract line counts, theorem and \texttt{def} counts, residual  \texttt{sorry} and axioms. Finally, conditioning on these metrics, we assign a binary easy/hard label using \texttt{claude-haiku-4-5} as judge, together with
a confidence score and a short rationale.\footnote{We confirmed judgment against real \texttt{claude-sonnet-4-6} performance, and found approx. 70\% correct predictions.}

\emph{Side effects.}
\label{sec:threats-axioms} A particular design issue is how to handle side effect in code.
PBTs in our corpus come from messy real-world environments: production code routinely touches databases, sockets, the filesystem, the wall clock, webhook queues, and so forth. 
The formalization agent handles this by treating the effectful world as an \emph{uninterpreted interface}. Parts of the source essential to the property---the entities and operators it constrains---are translated into Lean structures and inductive types. Parts that are pure plumbing---the database, the queue, the system clock---are introduced as opaque \texttt{axiom} declarations and threaded through the theorem as parameters. The resulting theorem holds \emph{over all interpretations} of those axioms, which is exactly the contract the original PBT was checking against the live environment.

\section{Results}
\Secl{results}

Our pipeline for the \FVSpec:FV dataset yields
\textbf{9{,}415 samples} drawn from \textbf{2{,}772 distinct PBTs}:
formalization succeeded for $2{,}623$ of $7{,}413$ distinct PBTs ($35\%$; Figure~\ref{fig:pipeline}, bottom right); a further 149 formalized PBTs are not traceable to the current public release. The mean sample carries 8 theorems
(median 6, max 66) and median structural faithfulness 0.65 (Figure~\ref{fig:structural_faithfulness}). A complete Lean implementation was generated for 59\% of samples; the remaining 41\% ship
against a stub \texttt{Impl.lean}, so the theorem statement is well-typed
but the body it constrains is opaque. The Claude Haiku grader splits the
dataset 62\%/38\% hard/easy. Where a PBT yields multiple compiling
formalizations, all are kept; the
\texttt{is\_canonical} field marks the highest-quality version per PBT.

The pipeline formalizes each of the 7{,}850 raw PBT records independently. Deduplicating those records onto 2{,}772 unique PBTs by content (\Secr{dataset-dedup}) collapses the PBTs but not their formalizations, leaving a mean of 3.4 formalizations per PBT (range 1--15). Among PBTs with multiple formalizations, 32\% have no single dominating attempt: for each formalization in the group there exists another that scores higher on at least one sub-metric of our faithfulness metric. In practice, this means for example, that the attempt with the best assertion coverage may have worse parameter coverage than a sibling; keeping all formalizations and flagging the best one with \texttt{is\_canonical} therefore captures strictly more information than retaining only the top-scoring attempt.

\subsection{Translation Quality}\label{sec:translation-quality}

\begin{figure*}[t]
  \centering
  \begin{minipage}[b]{0.52\linewidth}
    \centering
    \includegraphics[width=\linewidth]{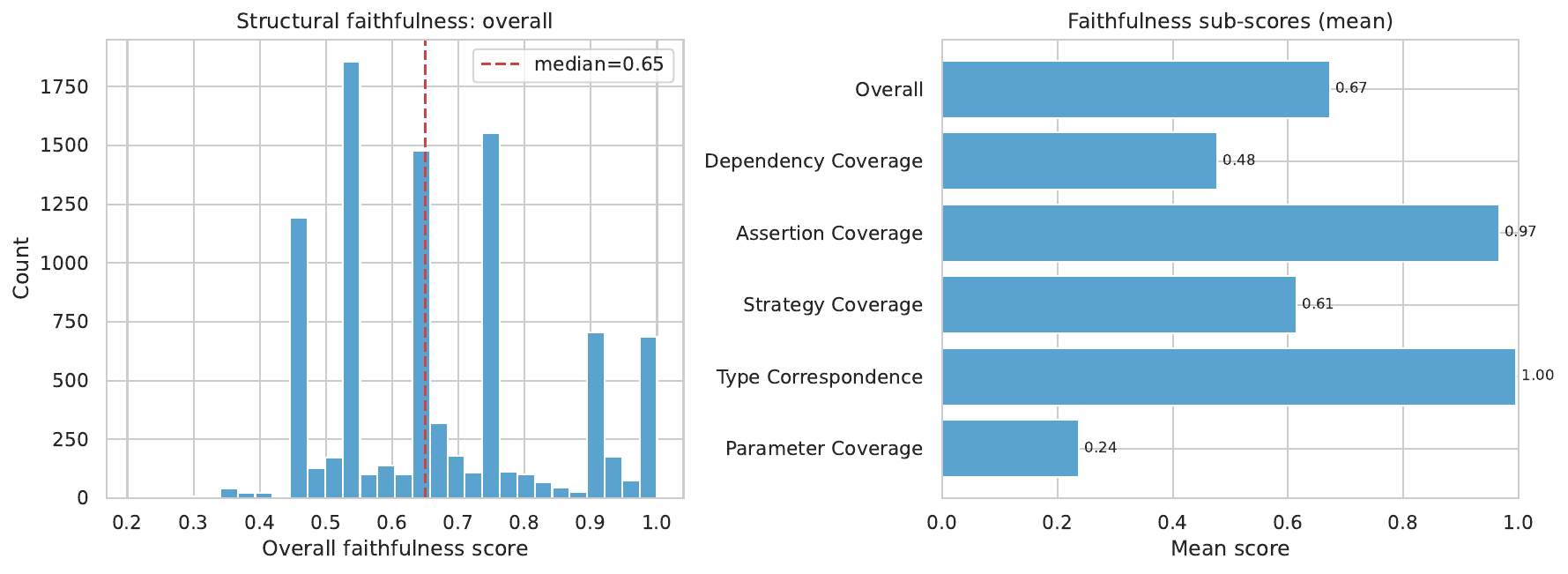}\\[2pt]
    \small (a) Structural faithfulness
  \end{minipage}\hfill
  \begin{minipage}[b]{0.48\linewidth}
    \centering
    \includegraphics[width=\linewidth]{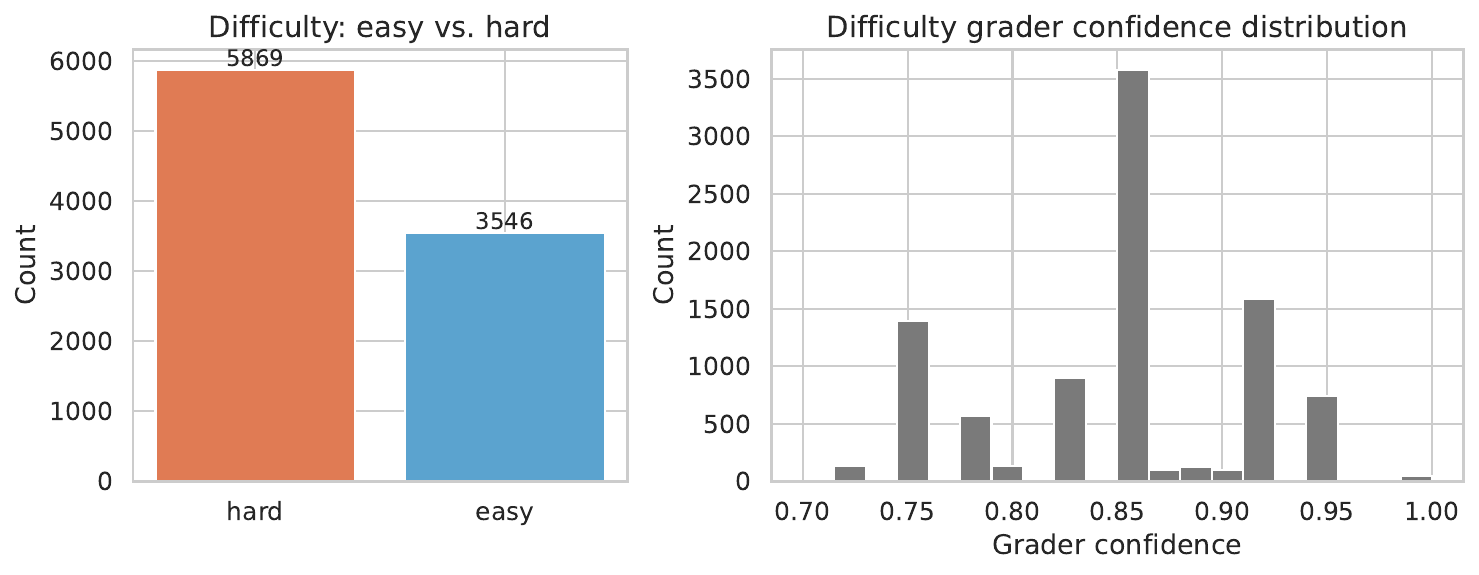}\\[2pt]
    \small (b) Difficulty grading
  \end{minipage}
  \caption{\textbf{(a)}~Structural faithfulness: distribution of the composite score (left) and mean sub-scores (right). \textbf{(b)}~Difficulty grading: easy vs.\ hard counts (left) and grader confidence distribution (right).}
  \label{fig:structural_faithfulness}
  \label{fig:difficulty_distribution}
\end{figure*}

We evaluate translation quality using \emph{structural faithfulness}. For this, we measure five sub-scores (parameter coverage, type correspondence, strategy coverage, assertion coverage, and dependency coverage) for both the original code and the translation, and we compute a weighted average of the difference. Figure~\ref{fig:structural_faithfulness} shows the distribution of the overall score and the mean of each sub-score across the dataset. The overall faithfulness distribution exhibits a clear mode above 0.5, indicating that the majority of translations preserve the essential structure of the source PBT, though a long tail of lower-quality translations remains.

\begin{figure*}[t]
  \centering
  \includegraphics[width=0.8\linewidth]{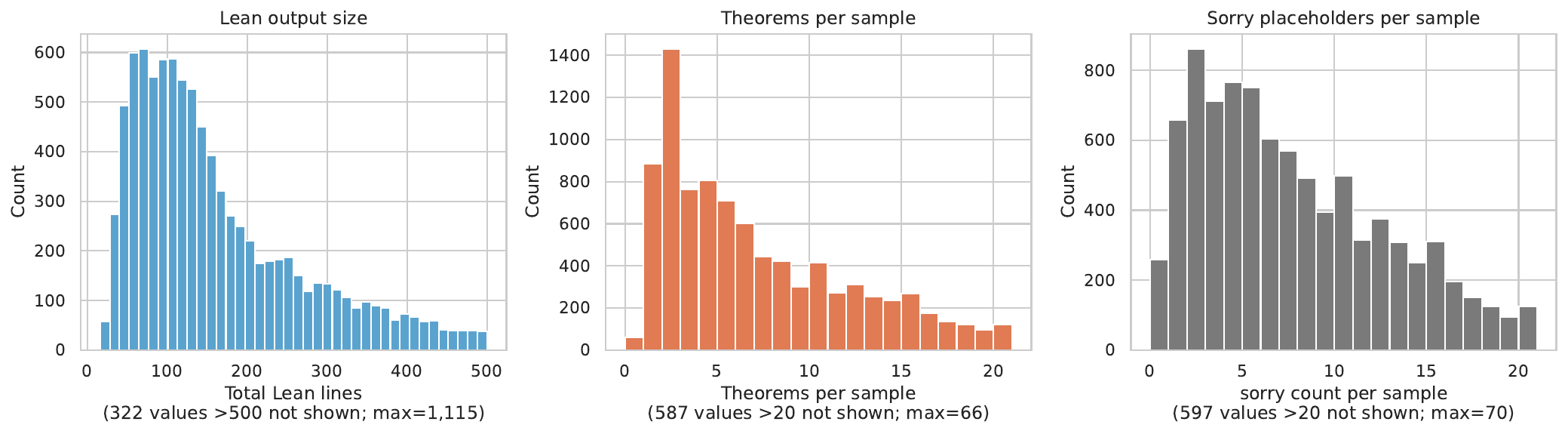}
  \caption{Total lines (left), theorems per sample (center), and \texttt{sorry} placeholders per sample (right).}
  \label{fig:lean_complexity}
\end{figure*}

In 59\% of samples the agent generated a complete Lean implementation (\texttt{Impl.lean} contains a computable definition); the remaining 41\% use a signature-only placeholder. Structural faithfulness is comparable across both groups (mean 0.67 vs.\ 0.68), confirming that specification quality is largely independent of implementation recovery.
The generated Lean artifacts span a wide range of complexity (Figure~\ref{fig:lean_complexity}): the median contains multiple theorems, and the \texttt{sorry} count tracks the theorem count closely.

\subsection{Difficulty Grading}

Figure~\ref{fig:difficulty_distribution} shows the results of our difficulty grading and the distribution of grader confidence. The benchmark contains a substantial proportion of both easy and hard problems, and the grader expresses high confidence for the majority of assessments. Difficulty is estimated here as a \textit{prediction}, where we ask the grader (haiku) to predict whether or not an agent would be able to complete the proofs. 

\begin{figure*}[t]
  \centering
  \includegraphics[width=0.7\linewidth]{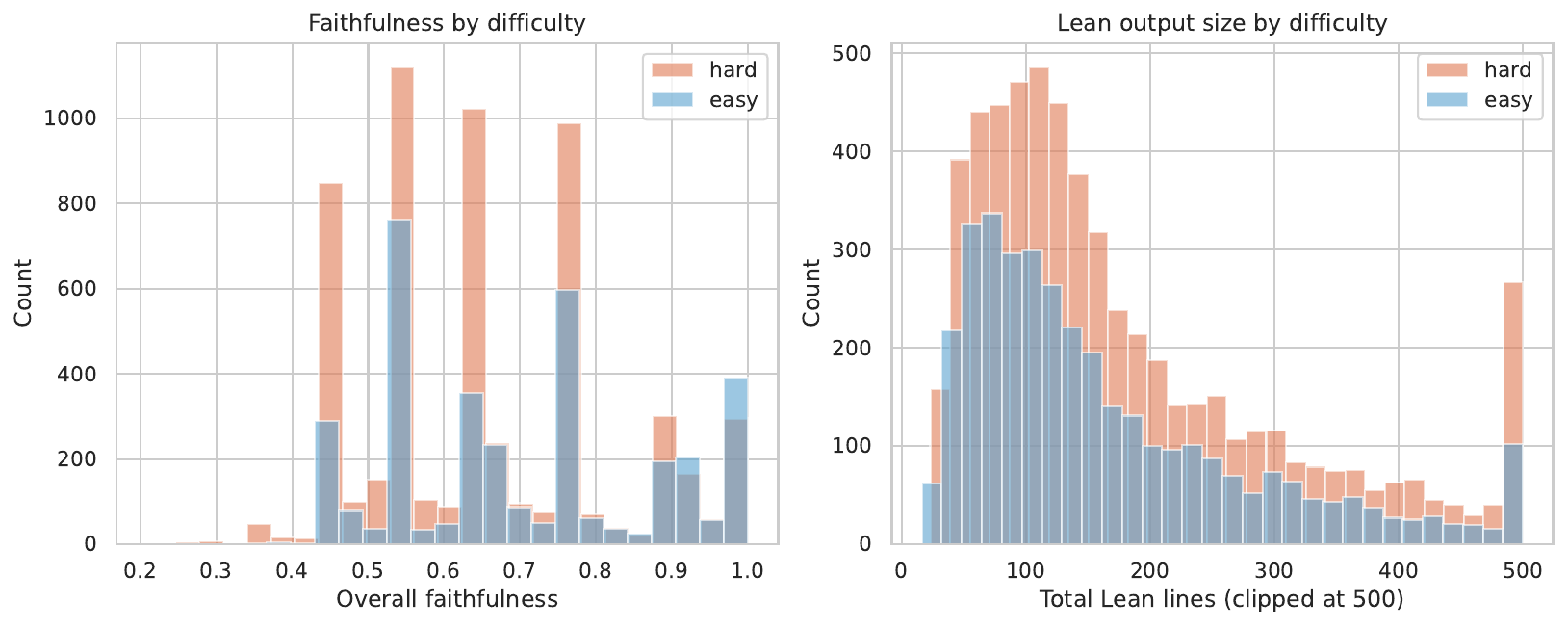}
  \caption{Overall faithfulness (left) and total Lean lines (right), stratified by difficulty.}
  \label{fig:difficulty_vs_faithfulness}
\end{figure*}

Figure~\ref{fig:difficulty_vs_faithfulness} examines the relationship between difficulty and two output characteristics: structural faithfulness and Lean output size. Hard problems tend to produce longer Lean outputs, consistent with the intuition that more complex source PBTs lead to more elaborate formalizations.

\subsection{Baselines}

\begin{figure*}[t]
  \includegraphics[width=\linewidth]{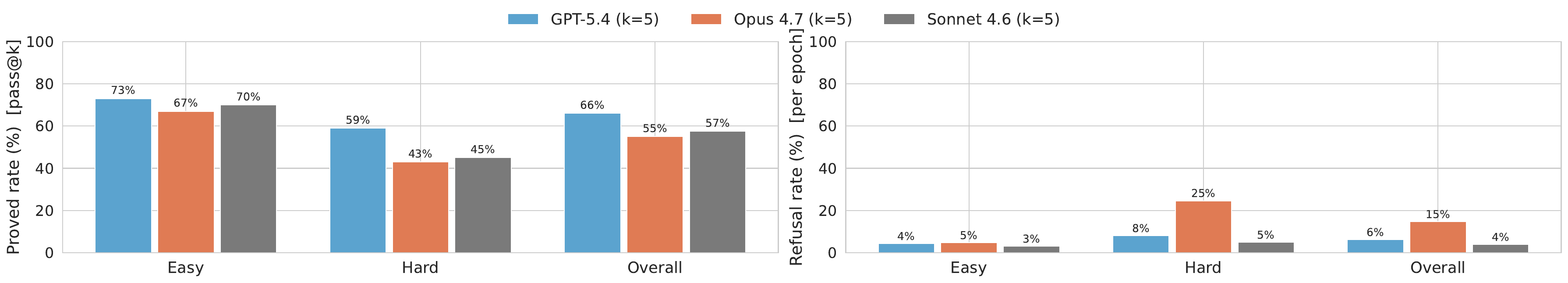}
  \caption{Left: proved-ever rate by difficulty (best-of-k); right: baseline rate of the model giving up.}
  \label{fig:baseline_prove_rate}
\end{figure*}

We evaluate three frontier models (Claude Sonnet~4.6, Opus~4.7, and GPT~5.4) on 100 randomly sampled easy problems and 100 hard problems. Each model has access to the Lean LSP via MCP tools and is scored on two metrics: a binary \emph{proved} flag (zero \texttt{sorry} remaining and \texttt{lake build} succeeds) and \emph{partial credit} (fraction of \texttt{sorry} placeholders removed, clamped to $[0,1]$).
Figure~\ref{fig:baseline_prove_rate} shows the models solved on average 49\% of hard samples and 70\% of easy ones.

\section{Limitations}
\Secl{threats}

Specifications are translated rather than hand-written, so some may not faithfully represent the original PBT. Our structural-faithfulness score is a heuristic: a high score does not guarantee semantic equivalence, and a low score does not necessarily indicate a useless problem.

This especially goes for our axiomatization approach to side effects, which has some specific limitations. If the property is fundamentally about timing, memory, ordering of side effects, or wall-clock latency, an uninterpreted axiom cannot witness it. \texttt{assert response\_time < 100ms} cannot become a meaningful Lean theorem because the relevant semantics is precisely what the axiomatization discards. Also, the agent identifies effectful boundaries via type signatures and call patterns. When a function has a pure signature but performs side effects internally (e.g., a logger called via a global, an \texttt{lru\_cache}-style hidden state), the agent will inline it as if pure. The resulting theorem may compile and even be true, but it is not guaranteed that its proof obligation always corresponds to what the original's code would have been.

Some formalizations are unfaithful in ways the score does not catch. Sample~30 (\texttt{pbt\_id}~990), \texttt{test\_alert\_collection\_holds\_unrelated\_alerts} from \texttt{bjoern-reetz/cap-tools}, has faithfulness 0.75 and is graded easy, and its theorem is true, but it misstates the property. \texttt{AlertState} is a dictionary keyed by each alert's \texttt{(sender, identifier, sent)} reference, so ingesting an existing reference overwrites that entry, and \texttt{UPDATE} and \texttt{CANCEL} messages delete entries. The Lean version is an append-only list of contentless alerts, and the theorem says that folding \texttt{ingest} over a list returns that list. The agent saw neither the real \texttt{AlertState}, because dependency resolution found nothing for this test, nor the precondition that references never collide, which the test's Hypothesis strategy enforces outside the test body. A faithful statement needs a keyed collection and an explicit \texttt{Nodup} hypothesis on the references. Dependency coverage, one component of the score, is 1.0 whenever no dependencies are resolved, which is the case for 3{,}793 samples (40.3\%).

Our difficulty labels come from \texttt{claude-haiku-4-5}, itself a model whose judgments are imperfect and uncalibrated against human expert performance. We expose the grader's confidence and rationale per sample so consumers can audit or override the labels.

A complete Lean implementation is generated for 59\% of samples; the remaining 41\% ship against a stub, weakening the proof obligation: the theorem is well-typed but the body it constrains is opaque. These samples are tagged for easy filtering.

Our scraper skips repositories with explicitly non-permissive licenses but accepts repositories where GitHub cannot detect a license at all.

We do not measure human expert performance on \FVSpec problems and therefore cannot calibrate the dataset's difficulty against the proof effort required of skilled humans.

\section{Related Work}
\Secl{related}
Our work extends FVAPPS \citep{dougherty2025proving}, which established AI-driven translation of coding problems into Lean. In FVAPPS, programs were taken from the APPS benchmark \citep{hendrycks2021apps}, then specs were inferred and implemented as PBTs, then transpiled to Lean theorem statements. We apply the same process at scale to PBTs found in the wild, inferring specifications from natural language and source code.
We enumerate related works below, and in Table~\ref{tab:fv-benchmarks}.

\emph{Lean verification benchmarks.}
A growing number of benchmarks target Lean. \citet{brandfonbrener2024minicodeprops} provide 177 program specs translated from Tons of Inductive Programs, finding that current LLMs  fail on medium and hard properties. \citet{thakur2025clever} offer 161 curated problems sourced from HumanEval, testing end-to-end verified code generation. \citet{wu2025verina} cover all three foundational tasks---code generation, specification generation, and proof generation---across 189 manually curated coding tasks. \citet{stechly2025veribench} translate Python code with documentation into Lean implementations, specifications, and unit tests (113 tasks). \citet{bursuc2025benchmark} support Lean, Rust/Verus, and Dafny, reporting success rates from 27\% (Lean) to 82\% (Dafny). \citet{ma2025veriequivbench} use LeetCode transformations to evaluate verifiable agents on novel data without contamination. These benchmarks are valuable but share a common limitation: their specifications derive from synthetic coding problems or pedagogical exercises, not from properties written by practicing engineers.

\emph{Other verification tool benchmarks.}
\citet{loughridge2024dafnybench} provide Dafny verification tasks.
\citet{beyer2025svcomp} provide the standard competition for software verifiers, with over 33{,}000 C tasks, but target automated verification rather than interactive theorem proving\footnote{These are completely different kinds of FV.}. \citet{wu2025invbench} focus specifically on loop invariant synthesis. \citet{chakraborty2024neural} explore neural synthesis for SMT-assisted proof-oriented programming in F*.

\emph{Interactive theorem proving benchmarks.}
\citet{hsiang2025leandojo} provide 122{,}517 theorems from Mathlib4. Similarly, \citet{thompson2024rango} collect 196{,}929 theorems from 2{,}226 GitHub Coq repositories. These are primarily mathematical rather than software-oriented, and their theorems exist in the training data of most frontier models.

\emph{Mathematics benchmarks.}
\citet{glazer2024frontiermath} evaluatet advanced mathematical reasoning. Although FV and mathematics share proof infrastructure, the skills required differ substantially: software verification demands modeling of program semantics, library APIs, and computational behavior, rather than abstract mathematical reasoning.

\FVSpec differs from each existing benchmark in at least one of the following three ways. First, our specifications originate from real-world property-based tests written by practicing engineers, not synthetic problems or pedagogical exercises. Second, the source PBTs were never formally verified, so most theorems in our benchmark will not appear in any model's training data. Third, we operate at a larger scale than existing Lean verification benchmarks, with thousands rather than hundreds of problems. Our work is motivated by the broader context of AI safety proposals premised on AI-driven formal verification.

\section{Conclusion}
\Secl{conclusion}
We present \FVSpec, a benchmark of 9{,}415 Lean~4 formal verification challenges derived from 7{,}413 distinct real-world Python property-based tests scraped from 497 open-source repositories. Unlike prior FV benchmarks---which draw from expert-written ITP code, competitive programming problems, or curated verification exercises---\FVSpec sources its specifications directly from practicing engineers who had no formal verification intent, placing its problems genuinely out of distribution relative to models' training data. Our agentic formalization pipeline produced 9{,}415 challenges from 2{,}623 of those PBTs (35\%), with median structural faithfulness 0.65 and 62\% of challenges classified as hard by a difficulty predictor. Baseline evaluations of three frontier models confirm that the benchmark is far from saturated, particularly on hard problems, leaving substantial room for progress on AI-assisted formal verification of ordinary software. Our approach, which we have described in depth, could easily be extended to PBTs from other languages (Haskell/QuickCheck, C++/RapidCheck, TypeScript/fast-check) and other ITPs (Coq, Isabelle, F$^\star$, Verus). Other potential future work includes merging with the HC-2026 corpus to add runtime context, and using disproved properties as seeds for program-repair challenges---bugs the original developer's test harness missed, now made precise in Lean.

\bibliographystyle{abbrvnat}
\bibliography{citations}

\newpage

\appendix

\section{Overview of other datasets}
\begin{table}[h]
\centering
\small
\setlength{\tabcolsep}{4pt}
\renewcommand{\arraystretch}{1.05}
\begin{tabularx}{\textwidth}{@{}
    >{\raggedright\arraybackslash}p{5cm}
    l
    r
    >{\raggedright\arraybackslash}X@{}}
\textbf{Benchmark} & \textbf{Language} & \textbf{Theorems} & \textbf{Focus} \\
\midrule
Minif2f~\cite{zheng2021minif2f}
  & Lean, Isabelle, HOL Light, Metamath & 488 & math Olympiad \\
PutnamBench~\cite{tsoukalas2024putnambench}
  & Lean, Isabelle, Coq/Rocq & 640 & math Olympiad \\
Fimo~\cite{liu2023fimo}
  & Lean & 149 & math Olympiad \\
DafnyBench~\cite{loughridge2024dafnybench}
  & Dafny & 782 & scraped Dafny proofs \\
ProofNet~\cite{azerbayev2023proofnet}
  & Lean & 371 & U math \\
RLM25~\cite{poiroux2025reliable}
  & Lean & 619 & G math \\
Lean Workbook~\cite{ying2024lean}
  & Lean & 57{,}000 & HS, U, G math \\
FormalMATH~\cite{yu2025formalmath}
  & Lean & 5{,}560 & math Olympiad \\
LEAN-GitHub~\cite{wu2024lean}
  & Lean & 28{,}597 & scraped Lean proofs \\
ProverBench~\cite{ren2025deepseek}
  & Lean & 325 & math Olympiad \\
MiniCTX~\cite{hu2024minictx}
  & Lean & 762 & six Lean projects \\
LeanGeo~\cite{leangeo}
  & Lean & 122 & HS, U, G math \\
IneqMath~\cite{lu2025solving}
  & Lean & 1{,}522 & math Olympiad \\
FATE-\{M,H,X\}~\cite{jiang2025fate}
  & Lean & 350 & U, G math \\
VeriBench~\cite{barkallah2025veribench}
  & Lean & 857 & software verification \\
FVAPPS~\cite{dougherty2025proving}
  & Lean & 17{,}931  & software verification (from APPS) \\
CLEVER~\cite{thakur2025clever}
  & Lean & 161 & software synthesis \& verification \\
Vericoding~\cite{bursuc2025benchmark}
  & Lean, Dafny, Verus & 12{,}504 & software synthesis \& verification \\
MUSTARD~\cite{huang2024mustard}
  & Lean & 5{,}866 & HS, U math \\
FormL4~\cite{lu2024process}
  & Lean & 14{,}510 & U math \\
CoqStoq (Rango)~\cite{thompson2024rango}
  & Coq/Rocq & 10{,}396 & software verification \\
MSC-180~\cite{li2025msc}
  & Lean & 180 & U, G math \\
HOList~\cite{bansal2019holist}
  & HOL Light & 29{,}462 & U, G math \\
FVSpec:FV & Lean & 75{,}005 & software verification: specs from the wild \\ 
\bottomrule
\end{tabularx}
\caption{\textbf{Prior ITP benchmarks.} HS = high-school, U = undergraduate, G = graduate. If one benchmark improves upon another, we exclude the lesser.}
\label{tab:fv-benchmarks}
\end{table}

Find an overview of prior ITP benchmarks in Table~\ref{tab:fv-benchmarks}.

\section{Comparison with the Hypothesis Corpus}
\label{sec:dataset-hc}

We released our corpus on HuggingFace in January 2026. In April 2026, while we were finalizing this work, the Hypothesis maintainers released their own scraped corpus~\cite{devoe2026hypothesis} (HC-2026).
The two corpora target different downstream uses---HC-2026 is built to study Hypothesis \emph{runtime} behavior (entropy consumption, predicate firing, coverage), while ours is built to feed a Lean formalization pipeline---which drives the design differences summarized in Table~\ref{tab:dataset-comparison}.

A consequence of HC-2026's runtime focus is that it filters out as ``invalid'' any repository whose PBTs may not be directly runnable without some custom setup: code that depends on external services, API keys, environment variables, or live databases. Our pipeline imposes no such requirement, since we only need a PBT's source code to translate it into Lean. Of our 497 repositories, 67 (13\%) appear in HC-2026's valid (executable) set and 195 (39\%) appear in the pre-validity superset. HC-2026 rejected most of the other 128 because it found no Hypothesis tests after installing them (80) or judged them near-duplicates of another repository (28).


\begin{table}
\centering
\begin{minipage}[t]{0.36\textwidth}
  \centering
  \caption{PBT corpus statistics.}
  \label{tab:dataset-stats}
  \footnotesize
  \setlength{\tabcolsep}{3pt}
  \begin{tabular}{lr}
    \toprule
    \textbf{Metric} & \textbf{Value} \\
    \midrule
    PBTs (raw scrape)              & 54{,}345 \\
    PBTs (deduplicated)            & 8{,}314 \\
    PBTs (public release)          & 7{,}413 \\
    Unique repositories            & 497 \\
    Unique GitHub owners           & 431 \\
    Upstream project names         & 453 \\
    Top-1 / top-10 PBT share       & 11.5\% / 30.8\% \\
    \midrule
    \multicolumn{2}{@{}l}{\emph{Per-repo (min / med / max):}} \\
    PBTs per repo                  & 1 / 4 / 852 \\
    Stars                          & 0 / 0 / 81{,}185 \\
    Forks                          & 0 / 0 / 29{,}652 \\
    \midrule
    \multicolumn{2}{@{}l}{\emph{Per-PBT (min / med / max):}} \\
    PBT LoC                        & 2 / 10 / 174 \\
    Direct fn deps                 & 0 / 0 / 12 \\
    \bottomrule
  \end{tabular}
\end{minipage}\hfill
\begin{minipage}[t]{0.62\textwidth}
  \centering
  \setlength{\tabcolsep}{3pt}
  \caption{Comparison with HC-2026~\cite{devoe2026hypothesis}.}
  \label{tab:dataset-comparison}
  \footnotesize
  \begin{tabular}{@{}l c c@{}}
    \toprule
    & \textbf{\FVSpec:PBT} & \textbf{HC-2026} \\
    \midrule
    Discovery        & dep.\ graph        & GitHub search \\
    Raw PBTs         & 54{,}345           & 38{,}885 \\
    Tests (deduped)  & 7{,}413            & 28{,}928 \\
    Repos            & 497                & 1{,}529 \\
    Repo dedup       & none               & \texttt{is\_fork} \\
    Test dedup       & SHA-256            & parametrization \\
    License filter   & no copyleft        & none \\
    Per-test data    & source + metrics   & runtime + coverage \\
    \midrule
    Repo overlap (valid)  & \multicolumn{2}{c}{67 / 497 \enskip(13\%)} \\
    Repo overlap (any)    & \multicolumn{2}{c}{195 / 497 \enskip(39\%)} \\
    \bottomrule
  \end{tabular}
\end{minipage}
\end{table}


\end{document}